\documentclass[aps,eqsecnum,twocolumn,amsmath]{revtex4}
\usepackage{graphics}
\usepackage[letterpaper,dvips,width=7.5in,includemp=false]{geometry}

\begin{document}

\title[title]{The quantization of exotic states in $SU(3)$ soliton models: A solvable quantum mechanical analog}%
\author{Aleksey Cherman, Thomas D. Cohen, Abhinav Nellore}%

\affiliation{Department of Physics, University of Maryland,
College Park, MD 20742-4111}
%\email{placeholder}%

%\date{}%
%\dedicatory{}%
%\commby{}%
% ----------------------------------------------------------------
\begin{abstract}
The distinction between the rigid rotor and Callan-Klebanov
approaches to the quantization of $SU(3)$ solitons is considered in
the context of exotic baryons.  A numerically tractable quantum
mechanical analog system is introduced to test the reliability of
the two quantization schemes.  We find that in the equivalent of the
large $N_c$ limit of QCD, the Callan-Klebanov approach agrees with a
numerical solution of the quantum mechanical analog.  Rigid rotor
quantization generally does not.  The implications for exotic
baryons are briefly discussed.
\end{abstract}
\maketitle
% ----------------------------------------------------------------
\section{Introduction}

Recent experimental reports\cite{exp} of the observation of an
exotic $\theta^{+}$ baryon have rekindled interest in  the
quantization of exotic states in $SU(3)$ soliton
models\cite{Coh03}. The soliton models are unique among the
theoretical tools used to analyze these reported states because
they predate the experiment and predict the mass of the state
very closely to the claimed experimental values
\cite{Pres,DiaPetPoly}.

A commonly employed technique for quantizing solitons treats the
system as a rigid rotor in which the soliton collectively rotates,
with its internal degrees of freedom fixed from the classical
solution\cite{Guad,SU3Quant}.  This technique is clearly correct
for $SU(2)$ chiral solitons when treated in the large $N_c$ limit
\cite{ANW} (where $N_c$ is the number of colors in QCD). However,
the validity of applying the rigid rotor approach to the
quantization of $SU(3)$ solitons has been questioned in
situations where the Wess-Zumino term can play an active role in
the dynamics of the system --- as it does for exotic baryon states
\cite{Coh03,Coh04,IKOR,Pob}. This issue remains controversial
\cite{Coh03,IKOR, Pob,DP,Coh04}.  In this paper we study a
numerically tractable quantum mechanical system that has many
critical features in common with $SU(3)$ solitons to obtain
insight about the underlying issues.

It should be noted that there are {\it two}  major approaches in
the literature for quantizing $SU(3)$ solitons. One is the rigid
rotor approach mentioned above, in which the collective
rotational and vibrational modes of the soliton are assumed to be
decoupled, and only the rotational modes are
quantized\cite{SU3Quant}. This approach has a number of benefits:
it is relatively simple to implement and is a straightforward
generalization of the Adkins, Nappi and Witten procedure which is
known to be justified at large $N_c$ for the case of $SU(2)$
solitons\cite{ANW}. Moreover, it is known to be justified at
large $N_c$ for non-exotic collective states in $SU(3)$ models.
The vast majority of the published studies of exotic baryons in
the context of soliton models have used this approach. The other
approach to quantizing $SU(3)$ solitons is the Callan-Klebanov
approach \cite{CK,IKOR}. This scheme is most easily understood in
the case of broken $SU(3)$, in which case excitations carrying
strangeness are unambiguously vibrational states, and should be
quantized as harmonic vibrations. It has been argued that this
method remains valid as one approaches the $SU(3)$ limit
\cite{IKOR}. In fact, for the non-exotic states one of the
vibrational modes becomes softer and goes to zero frequency as the
$SU(3)$ limit is approached, and thus reproduces the results of
rigid rotor quantization. Therefore it seems that this method
should be valid for both broken and unbroken $SU(3)$. However,
for exotic states the Callan-Klebanov approach does {\it not}
reproduce the rigid rotor result; indeed when applied to the
original Skyrme model it gives {\it no} exotic resonant states at
all\cite{IKOR}. The Callan-Klebanov approach has the obvious
disadvantage of being more difficult to implement; this may
explain the fact that it has not been widely used in studies of
exotic baryon states.  Indeed, to the best of our knowledge only
two such calculations have been reported\cite{IKOR,PR}.

The fact that the Callan-Klebanov method gives different results
from rigid-rotor quantization for exotic states implies that at
least one of these methods is wrong. Assuming that one is correct,
it is critical to know which one.  It has been argued elsewhere
on a number of grounds that the rigid-rotor approach is the
culprit\cite{Coh03,IKOR,Coh04,Pob}, and the Callan-Klebanov method
is correct, at least at large $N_c$.  While there have been
attempts in the literature to rebut at least some of these
arguments \cite{DP}, it has been argued that these rebuttals are
fundamentally flawed\cite{Coh04}. We will not attempt to recap
these arguments but instead refer the reader to the original
literature.

Our purpose here is simply to consider a tractable quantum
mechanical system that has states analogous to the exotic states of
the $SU(3)$ soliton, which is analytically intractable.  This system
can be solved numerically (to essentially any desired degree of
accuracy) and via both approximate methods --- the Callan-Klebanov
approach and the rigid-rotor approach in the analog of the large
$N_c$ limit. One can then explicitly see which approach works.  As
we will show below, the Callan-Klebanov method reproduces the
numerical solutions for this model up to expected errors of order
$1/N_c$, while the rigid-rotor approach generally fails.

The model considered here was introduced in
refs.~\cite{Coh04,Pob}. In those works it was observed that the
Callan-Klebanov approach and rigid-rotor approach gave different
answers for the excitation spectrum of the model, and it was
argued on semiclassical grounds that the assumptions underlying
the rigid-rotor quantization were not self consistent.  However,
these works did not  show explicitly that the Callan-Klebanov
approach actually produces the correct spectrum, and that the
rigid rotor approach does not.

This paper is organized as follows.  In the next section, the model
will be introduced.  The two subsequent sections will implement the
rigid rotor quantization and the Callan-Klebanov quantization for
this model. (Some details of the Callan-Klebanov treatment are
relegated to an appendix). The next section contains a brief
discussion of the numerical solution of the model and a comparison
of the numerical solution with the two approximation methods.
Finally, we discuss the implications of our results for soliton
treatments of exotic baryons.

\section{A Tractable model}

We wish to study a numerically soluble model that incorporates the
relevant features of the excited states of $SU(3)$ solitons. Given
the nature of the critique of the rigid rotor treatment
\cite{Coh03,IKOR,Coh04,Pob}, in order to mimic the soliton
problem, we need a system in which there are both collective
rotational and vibrational degrees with the same quantum numbers.
There should also be a force on the system that is topological in
nature and velocity dependent to mimic the effect of the
Wess-Zumino term (which is topological and first order in time
and thus acts on velocities).

To do this, let us consider the problem of a composite charged
particle moving nonrelativistically on the surface of a sphere of
radius $R$, which has a magnetic monopole of strength $g$ at its
center.  It was observed long ago by Witten that the motion of a
charged particle on the surface of a sphere in the field of a
magnetic monopole is topological in essentially the same way as
the motion of chiral fields in the presence of the Wess-Zumino
term\cite{Wit1}. Indeed, the original rigid rotor quantization of
$SU(3)$ solitons by Guadagnini was explicitly done in analogy to
the monopole problem\cite{Guad}.  It is important that we consider
a composite system, {\it i.e.}, one with internal degrees of
freedom. The dynamics of the composite system are analogous to
the internal dynamics of the soliton and the key issues are
associated with the possible interplay of internal and collective
degrees of freedom.

To be concrete, we consider our composite particle as being made of
two point-like constituent particles\cite{Coh04,Pob}.   One
constituent is a charged particle (with charge $q$).  The other
constituent particle is electrically neutral.  We take both
constituent particles to have the same mass ($M$). The particles
interact via a nonsingular potential that binds the particles
together.   To ensure that the system is rotationally invariant, the
potential can depend only on the separation between the particles.
Since the particles are strongly bound, and thus spend most of their
time near the minimum of the potential, we can take the interaction
to be due to an approximately harmonic potential of spring constant
$k$. As noted above, the magnetic field due to the monopole serves
to provide the desired velocity-dependent topological force.  The
analog of the classical static soliton is simply the classical
configuration which minimizes the energy---namely, the two particles
on top of each other at the minimum of the potential. This
configuration will be referred to as the ``soliton''.

The  semiclassical treatment of the $SU(3)$ soliton is only
justified in the large $N_c$ limit of QCD.  Thus, it is important
that the various parameters in the toy model are chosen to scale
with $N_c$ in a manner that emulates the soliton case:
\begin{equation}\label{toyscale}
q \sim N_c^0 \; \; \;  R \sim N_c^0 \; \; \;  g \sim N_c^1 \; \;
\;
 M \sim N_c^1 \; \; \;  k \sim N_c^1 \; \; .
\end{equation}
These scaling rules ensure that energy of the classical ``soliton''
scales as $N_{c}^{1}$, the characteristic frequencies associated
with internal excitations of the ``soliton'' ($\sqrt{2k/M}$) scale
as $N_c^0$, and that the excitations associated with exotic motion
also scale as $N_c^0$.  This behavior is analogous to the $SU(3)$
soliton system \cite{Coh04}.

\section{Rigid-Rotor Quantization\label{RRQ}}

To develop some intuition about the rigid-rotor approach, first
consider a simpler problem: the charged composite particle moving
on the surface of a sphere {\it without} a magnetic monopole. Both
constituents of the particle have mass $M$ and interact via a
(nearly) harmonic potential with spring constant $k$.  The
classical ground state of the ``soliton'' has the two particles
on top of one another located at some point (say the north
pole).  Of course, since the ``soliton'' breaks rotational
symmetry,  this classical state is highly degenerate---the
``soliton'' can be localized at any point on the sphere.  Thus,
there exists a collective manifold of configurations, all with
the same classical ground state energy.

One can consider the system slowly moving through this
manifold---{\it i.e.}, the two particles move together in lock-step,
with the center of mass moving collectively around the sphere.
Without solving the quantum equations of motion one can see that
this approximation is justified quantum mechanically in the large
$N_c$ limit due to the large spring constant and the large mass in
Eq.~(\ref{toyscale}). To see this, first one assumes that the
intrinsic vibrations are decoupled from the collective motion, and
subsequently checks for self consistency. The characteristic energy
scale for low-lying collective excitations is just the inverse of
the moment of inertia $E_{\rm rot} \sim \frac{1}{2 M R^{2}} \sim
1/N_c $.  This is very small compared to the characteristic energy
associated with the intrinsic vibrations $E_{\rm vib} \sim  \sqrt{2
k/M} \sim N_c^0$. The fundamentally different scales at large $N_c$
allow the collective rotational motion to be essentially decoupled
from the intrinsic vibrational motion.   Thus the composite
system---our ``soliton''---behaves as if it were a single charged
particle of mass $2 M$ at large $N_{c}$. Indeed this is hardly
surprising: the strong spring constant ensures that in the large
$N_c$ limit the two particles are tightly bound, and thus move
together collectively. This is rigid rotor quantization since the
internal structure of the ``soliton'' is approximated as being rigid
and corrections to this are higher order in $1/N_c$.

Now consider what happens when the monopole field is added to the
system.  The spring constant remains strong and the wave function
for the composite system remains highly localized.  Thus it is
plausible that the two particles continue to move collectively
together in the same way that they did in the absence of the
monopole.  This reasoning suggests that the collective excitations
will be those of a single particle of mass of $2 M$ and charge $q$
moving in the field of the monopole.

To obtain the collective energy spectrum, we need only consider the
dynamics of a single charged particle (of mass $2M$ and charge $q$)
on a sphere of radius $R$ with a magnetic monopole of strength $g$
at its center. This system is well described in refs.
\cite{Wit1,Guad}. Any point on a sphere can be labeled by an element
of $SU(2)$.  (Technically, a point on a sphere $S^2$ corresponds to
a fiber in $SU(2)$; elements in a fiber are related by unitary
phases.) Thus the states of a single particle on a sphere correspond
to the irreducible representations of $SU(2)$, and these can be
written in terms of Wigner D-functions $D^{J}_{m,m'}$. Without a
magnetic monopole, a particle sitting on a sphere has no intrinsic
angular momentum and the $m'$ of the Wigner D-function is always
zero. The allowed states are then simply the spherical harmonics
$Y^{J}_{m}$. However, the presence of a magnetic monopole gives the
particle an intrinsic angular momentum of $q g$.  The reason for
this is simple.  A system which has both an electric and magnetic
field has momentum carried in the fields. A calculation of
$\vec{J}_{\rm field} = \vec{r} \times \vec{p}_{\rm field}$ for a
static classical field configuration yields  $\vec{J} = q g
\hat{r}$---the angular momentum along the intrinsic $z$ axis is $q
g$.  This restricts the allowed representations to those where $J
\geq q g$. Thus the states of the particle can be written as Wigner
D-functions $D^{J}_{m, q g}$, with $J \geq  q g$.

It is easy to find the energy of the system.  At the classical
level, the Hamiltonian is given by
\begin{equation}
H = \frac{J^2 - (q g)^2}{2 I}  \; .\label{Hclass}
\end{equation}
where $I$ is the moment of inertia.   The energy is purely kinetic
and is zero when the system is at rest, namely, for  $J = g q$.
For the present system $I = 2 M R^2$ where the factor of two
reflects the fact that the mass of the composite is $2 M$.  To
quantize the system one simply promotes $J$ to a quantum
operator.  The energy of the system is then given by
\begin{equation}
E(J) = \frac{J (J+1) - (q g)^{2}}{4 M R^2} \; \; {\rm with}\; \; J
\ge q g \; . \label {ERR}
\end{equation}
The excitation energy of the lowest-lying collective state is then
given by
\begin{equation}
\Delta E_{R} \equiv {E(q g+1) - E(q g)} = \frac{q g}{2 M R^2} \; .
\label{DeltaER}
\end{equation}

\section{Callan-Klebanov quantization}
\begin{figure}[t]
\begin{minipage}{2.5in}
\includegraphics{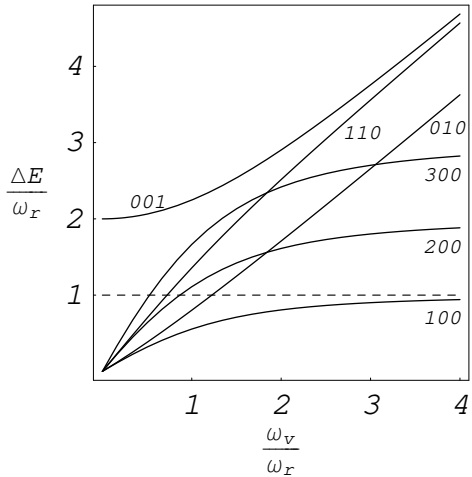} \label{anFig}
\caption{Semiclassical solutions of the system based on the
Callan-Klebanov approach as given in \cite{Coh04}. The energy is
measured in units of $\omega_r$ and the solutions are plotted as a
function of the ratio $\omega_v / \omega_r$, with $\omega_r =
\frac{q g}{2 M R^2}$ and $\omega_v = \sqrt{\frac{2 k}{M}}$. The plot
shows $\Delta E = E - E_{ground}$. Each solution is labeled by the
$n_1 n_2 n_3$ of Eq.~(\ref{n1n2n3}), giving its decomposition in
terms of the three non-zero positive normal modes of the system.}
\end{minipage}
\end{figure}

There is another way to look at the problem. First, consider the
case of a single charged particle moving on a sphere in the field
of a strong magnetic monopole from a classical perspective.   The
system is a charged particle moving in a strong magnetic field.
The paths of particles moving in magnetic fields bend, and in
strong fields they bend into classical orbits of small radius.
Thus, the particle does not make great circle orbits around the
sphere but instead makes tight local orbits. Moreover, it is well
known that the quantization of these localized orbits leads to
Landau levels---namely, the excitation spectrum of a harmonic
oscillator \cite{Landau}. Now consider the classical dynamics of
the complete toy model, with two interacting particles on a
sphere in the presence of a magnetic monopole.  There are two
types of harmonic dynamics: the orbits associated with motion of
the center of mass in the magnetic field, and the motion
associated with excitations of the two particles relative to one
another. These two types of motion each have characteristic
frequencies associated with them. The orbits of the center of
mass due to the magnetic field have a characteristic frequency,
\begin{equation}
\omega_r \equiv \frac{q g}{2 M R^2} \label{omegar} \; ,
\end{equation}
which is  the cyclotron frequency of a particle of mass $2M$ and
charge $q$ moving in the magnetic field of a monopole of strength
$g$ a distance $R$ away.  It is worth observing that $\omega_r$ is
precisely equal to the excitation energy in the rigid rotor
approximation of Eq.~(\ref{DeltaER}). The characteristic frequency
of the intrinsic vibrations is
\begin{equation}
\omega_{v} \equiv \sqrt{ \frac{2 k}{M} }  \; . \label{omegav}
\end{equation}
The key point is that the classical equations of motion can induce
mixing between these two types of motion.

The classical equations can be truncated at harmonic order and
then solved for the normal mode frequencies. Due to the presence
of velocity-dependent forces, it is useful to formulate the
problem in terms of coupled first-order differential equations for
positions and velocities.  This classical problem was analyzed in
refs.~\cite{Coh04,Pob} and this analysis is briefly recapitulated
in the appendix.  The problem has four degrees of freedom and
hence there are four normal mode frequencies (which come paired
as positive and negative frequency solutions to the equations of
motion).  Three of the normal mode frequencies are nonzero and
the fourth is a zero mode. The Callan-Klebanov approach amounts
to the quantization of these harmonic modes.

The zero mode is non-dynamical in nature: it corresponds to
relocating the position of the ``soliton'' to a new point on the
sphere but has no velocity associated with it.  Quantum mechanically
this mode corresponds to the nonexotic states of the system.  For
the present case it represents the ``excitation'' of one of the $2
qg +1$ degenerate ground states.  In the context of the $SU(3)$
soliton case it corresponds to the excitation of the usual hyperons
from the nucleon (which, of course, are exactly degenerate in the
SU(3) limit).

The non-zero modes correspond to physical harmonic excitations.  The
excitation spectrum to leading order in the $1/N_c$ expansion is
thus given by
\begin{equation}\label{n1n2n3}
\Delta E= E - E_{ground} = n_1 \omega_1 + n_2 \omega_2 + n_3
\omega_3
\end{equation}
where the $\omega_i$ are the three non-zero positive
eigenfrequencies. In Fig. 1, we plot a few representative
low-lying states. In this figure $\Delta E$ is given in units of
$\omega_r$ and it is plotted as a function of differing spring
constants which are reflected in the dimensionless ratio
$\frac{\omega_v}{\omega_r} = \frac{R^2 \sqrt{8 M k}}{q g}$.  The
advantages of working with these two dimensionless ratios should
be clear; the results only depend on particular combinations of
the parameters greatly simplifying the analysis.  Thus, it is
sufficient to work with fixed values of $M$, $R$, $q$ and $g$
while varying $k$ to explore all of the physics at large $N_c$.
It is worth noting that $\frac{\omega_v}{\omega_r}$ is of order
$N_c^0$ and thus the large $N_c$ limit can be taken for any value
of this ratio.

For the sake of comparison, in Fig.~1 we also plot the excitation
energy of the collective state predicted by the rigid rotor
quantization as a dashed line (located  at $\frac{\Delta
E}{\omega_r} = 1$) . One key observation needs to be made at this
point.  Clearly these two approaches predict different excitation
spectra for generic values of $\frac{\omega_v}{\omega_r}$.  The two
approaches do agree in the limit $\frac{\omega_v}{\omega_r}
\rightarrow \infty$ where $\omega_1 \rightarrow 1$, but as noted
above nothing in the large $N_c$ scaling rules tells us that this
ratio should be large.

It is apparent from this analysis that there are two conflicting
pictures for quantization in this problem.  At this stage it is
worth noting that the rigid-rotor approach describes far fewer
states in the spectrum than the Callan-Klebanov approach does. It
is important to determine which of these two quantization schemes
actually describes this system as the large $N_c$ limit is
approached.

\section{Numerical solution of the model}

To numerically solve the quantum system one must specify the
Hamiltonian in a particular basis. As shown in ref.~\cite{Guad},
Wigner D-functions with $J \geq eg$ form a basis of states for the
charged particle.  The states in this basis are labeled by $J$
($J \ge q g$), $m_J$ ($J \ge m_J \ge -J$) and ${m'}_J= q g$.   Of
course, the standard spherical harmonics form a basis of states
for the neutral particle and are labeled by $L$ and $m_L$ ($L \ge
m_L \ge -L$). We can obtain a basis of states for the composite
system by taking tensor products of the basis states for the
charged and neutral particles.

\begin{figure*}[t]
\label{numFig}
\begin{tabular}{ccc}
  \begin{minipage}{2in}
    \includegraphics{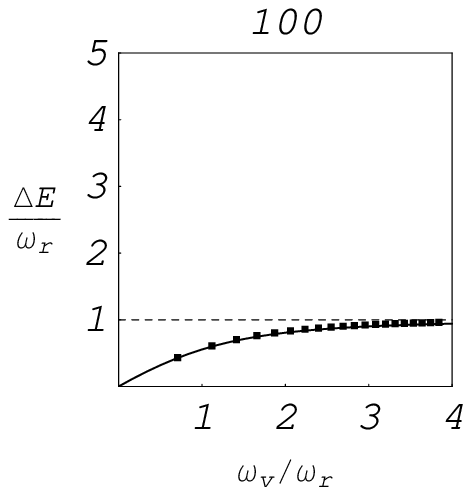}
    \end{minipage} &
    \begin{minipage}{2in}
        \includegraphics{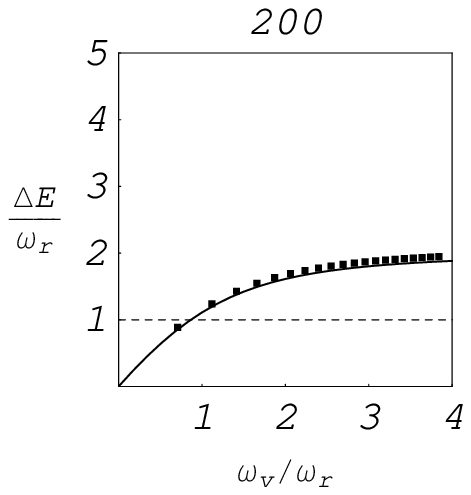}
    \end{minipage} &
    \begin{minipage}{2in}
        \includegraphics{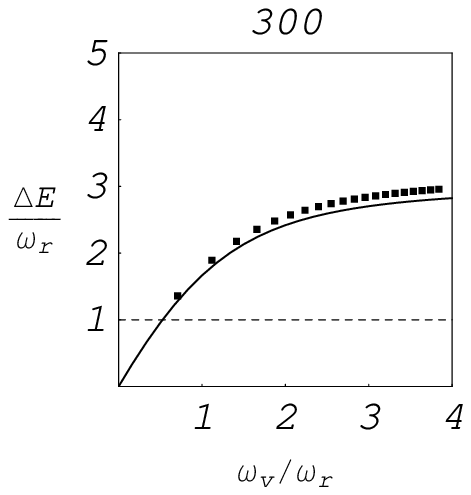}
    \end{minipage} \\
    \begin{minipage}{2in}
        \includegraphics{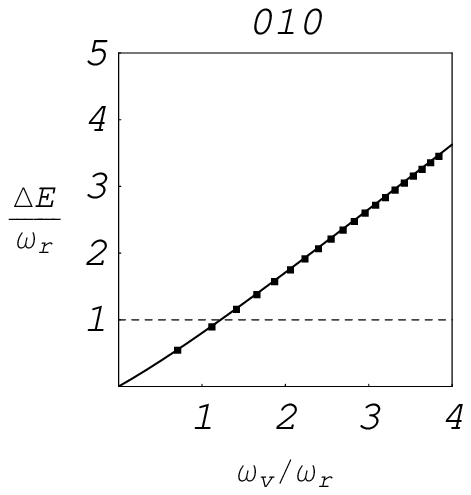}
    \end{minipage} &
    \begin{minipage}{2in}
        \includegraphics{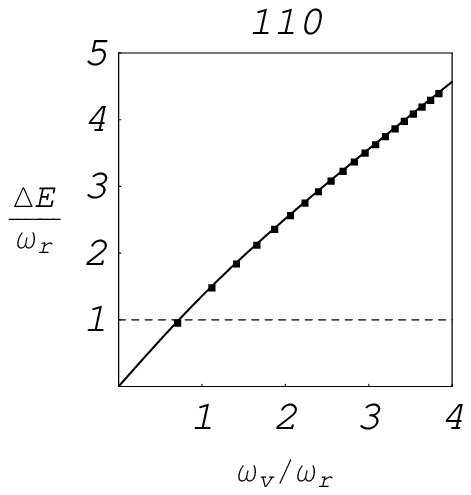}
    \end{minipage} &
    \begin{minipage}{2in}
        \includegraphics{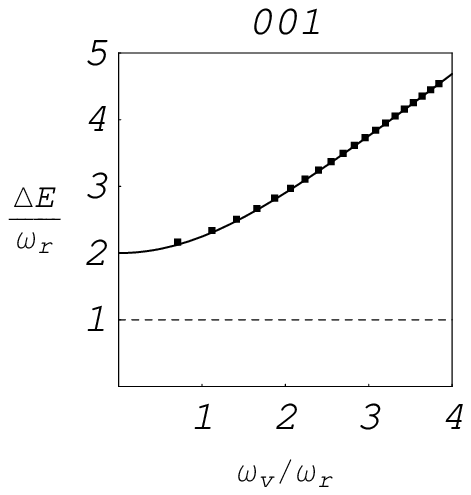}
    \end{minipage} \\
\end{tabular}
\caption{We plot the numerical simulation data (dots) against the
semiclassical approximation solutions based on the
Callan-Klebanov approach (solid lines) for each of the energy
levels shown in Fig.~1.  The plots are of ${\Delta}E = E -
E_{ground}$.  The numerical computation was done with $ e g = 40$
and $J_{max} = L_{max} = 48$.}
\end{figure*}

In the semiclassical analysis it is sufficient to consider a
harmonic potential in the large $N_c$ limit;  anharmonic effects
only come in as $1/N_c$ corrections.  Thus in the numerical model
studied, the interaction potential between the particles should be
approximately quadratic at short distances to mimic the soliton
problem. However, since the problem is posed on a sphere and we are
using angular variables, it is necessary that the potential be
periodic in the angular separation. The simplest form for the
potential with these properties is $k R^2 (1 - \cos{\gamma})$, where
$\gamma$ is the angular separation between the charged and neutral
particles.

The Hamiltonian for this system is  the sum of the kinetic
energies of the charged particle, the kinetic energy of the
neutral particles and the interaction potential between them:
\begin{equation}\label{H}
\hat{H}=\hat{H}_q+ \hat{H}_n+\hat{H}_{\rm int} \; .
\end{equation}
The charged particle kinetic energy is
\begin{equation}\label{Hq}
    \hat{H}_{q} = \frac{\hat{J}^2 \otimes \hat{\mathrm{1}}_n - (eg)^{2}}{2 M
    R^{2}} \; ,
\end{equation}
for the reasons discussed earlier.   The neutral particle term is
\begin{equation}\label{Hn}
    \hat{H}_{n} = \frac{ \hat{\mathrm{1}}_q \otimes \hat{L}^2}{2 M R^{2}}
\end{equation}
and the interaction term is
\begin{equation}\label{Hint}
    \hat{H}_{\rm int} = R^2 k (1 - \cos{\gamma}) \; .
\end{equation}

A straightforward computation using standard identities produces
the matrix elements of the total system Hamiltonian:
\begin{widetext}
\begin{eqnarray} \label{matElemEq}
&\langle J', {m'}_J, e g; L', {m'}_L|\hat{H}|J, m_J, e g; L, m_L
\rangle = \frac{J(J+1) +
L(L+1) - (eg)^{2}}{2 M R^{2}}\delta(J,J')\delta(m_J,-{m'}_J)\delta(L,L')\delta(m_L,-{m'}_L) + \nonumber \\
 & k R^2 \, \delta(J,J')\delta(m_J,-{m'}_J)\delta(L,L')\delta(m_L,-{m'}_L) - (-1)^{e g - {m'}_J +
 {m'}_L}\sqrt{(2J+1)(2J'+1)(2L+1)(2L'+1)}\times \nonumber \\
 &\sum_{c=-1}^{1}{
    \begin{pmatrix}
      J' & 1 & J \\
      - e g & 0 & e g \\
    \end{pmatrix}
    \begin{pmatrix}
      J' & 1 & J \\
      -{m'}_J & -c & m_J \\
    \end{pmatrix}
    \begin{pmatrix}
      L' & 1 & L \\
      0 & 0 & 0 \\
    \end{pmatrix}
    \begin{pmatrix}
      L' & 1 & L \\
      -{m'}_L & c & m_L \\
    \end{pmatrix}
 }
\end{eqnarray}
\end{widetext}
Here $\delta(i,j)$ is the Kronecker delta function, and the terms in
the summation are Wigner $3j$ symbols.  Armed with
Eq.~(\ref{matElemEq}), we can calculate the matrix of $\hat{H}$ in a
truncated basis (working up to some cutoff values of $J$ and $L$),
and find the lowest few eigenvalues.  Of course the system is
rotationally invariant, and hence the matrix block diagonalizes into
blocks of good total angular momentum and good z component of the
total angular momentum.  In principle, one can exploit this symmetry
to greatly reduce the size of the matrices considered.  It is quite
straightforward to exploit the third component of the total angular
momentum: to find the spectrum it is sufficient to study states of
total $m=0$ since all multiplets have an $m=0$ member.  Since the
$z$ component of the angular momentum is additive it is sufficient
to study basis states which have $m_J=-m_L$. Imposing good total
angular momentum is in principle straightforward, but in practice is
rather cumbersome due to the large number of terms in the
Clebsch-Gordan series (as a result of the large cutoffs needed for
convergence for the cases of numerical interest). Thus it was
simpler to work with the full matrices for total $m=0$. Taking $q
g=40$ to ensure large $N_c$, our matrices were of dimension
approximately $21,000$. Fortunately they are quite sparse and easily
amenable to sparse matrix techniques. With these methods the lowest
several eigenvalues of the system were easily calculated.

In Fig.~2, numerical solutions for several low-lying energy levels
are presented for the case $q g = 40$.  They are compared to the
semiclassical predictions based on the Callan-Klebanov approach; a
wide variety of spring constants are considered. The plots are
expressed in terms of the same dimensionless ratios as used in
Fig.~1. To keep the graphs uncluttered, we have presented each
energy level as a function of the strength of the spring constant
on a separate graph. We used the level ordering given in the
semiclassical expression, Eq.~(\ref{n1n2n3}), to associate
particular numerically computed energies with semiclassical
states; {\it i.e.}, the $n^{th}$ lowest numerical computed state
is associated with the $n^{th}$ lowest state in
Eq.~(\ref{n1n2n3}). There is some ambiguity with this method in
the immediate vicinity of level crossings in which case we used
numerical smoothness to determine the association of levels. It
is quite apparent that the actual energy levels of the system
closely follow the semiclassical treatment based on the
Callan-Klebanov approach. Moreover, generically they are rather
far from the prediction of rigid rotor quantization; namely, that
a collective state should exist at $\frac{\Delta E}{\omega_r} = 1$
(which is indicated in both Fig.~1 and Fig.~2 as a dashed line).

Of course, the semiclassical predictions based on the
Callan-Klebanov method are not expected to precisely reproduce the
spectra.  One expects $1/N_c$ corrections with a characteristic
scale of $(q g)^{-1}$ and that such corrections will increase with
increasing excitation energy due to anharmonicities.  Thus it seems
highly plausible that the small but discernible deviations of the
semiclassically predicted $2\, 0\, 0$ and $3\, 0\, 0$ modes from the
numerical simulation are due to such effects. Indeed it is easy to
see that the scale of these deviations is typical of what one
expects.  While it is generically nontrivial to compute the $1/N_c$
corrections, it is quite straightforward to do so in the limit
$\frac{\omega_v}{\omega_r} \rightarrow \infty$, which is the
infinitely strong coupling limit.  In this case, the dynamics
undoubtedly do reduce to that of a single particle of mass $2 M$,
and Eq.~(\ref{ERR}) holds for any $N_c$. This implies that in the
limit $\frac{\omega_v}{\omega_r} \rightarrow \infty$ the actual
value for $\frac{\Delta E}{\omega_r}$ in a state $n 0 0$ will exceed
the Callan-Klebanov prediction by an amount given by $\frac{n^2 +
1}{2 q g}$.  This is clearly a $1/N_c$ correction. It gives a
correction of $0.025$, $0.0625$ and $0.125$ for the $1\, 0\, 0$,
$2\, 0\, 0$ and $3\, 0\, 0$ states, respectively, in the strong
coupling limit. For the largest values of
$\frac{\omega_v}{\omega_r}$ we computed ($\frac{\omega_v}{\omega_r}
= 3.84$), the actual amount by which the Callan-Klebanov formula
underpredicted the numerical result was $0.0250$, $0.0739$, and
$0.147$, respectively, and it is highly plausible that these values
will asymptote to the known $1/N_c$ corrections in the strong
coupling limit. Thus, the size of the deviations from the
Callan-Klebanov predictions is of the scale expected from $1/N_c$
corrections.

Upon the completion of this work we learned of a calculation of the
spectrum of this model by Diakonov and Petrov.  Their results agree
with ours---the Callan-Klebanov results accurately describe the
spectrum while the rigid rotor does not\cite{private}.

\section{Conclusion}

The analysis of the toy system considered above shows explicitly
that for a system with topological velocity-dependent interactions
analogous to a dynamically active Wess-Zumino term, the
Callan-Klebanov method is the correct way to implement
semiclassical quantization.  On the other hand, rigid-rotor
quantization does not generally work in this system.

It is worth understanding why the plausible sounding argument given
in Sect.~\ref{RRQ} for rigid-rotor quantization fails. The key point
is that the intuition gained from the case without the
monopole---that the strong coupling present at large $N_c$ implies
that the center of mass of the system moves collectively---does not
automatically translate to the case where the monopole {\it is}
present. While the strength of the coupling remains large, the
effect of the monopole on the internal dynamics is also
parametrically large for large $N_c$.  The reason that the monopole
plays such a role should be clear.  If one imagines the center of
mass of the composite system slowly moving in an apparently
collective way, one finds that the monopole exerts a force only on
the charged constituent, but {\it not} on the neutral one. The
neutral constituent can only follow the charged one due to the force
exerted by the spring.  This implies possible mixing of the internal
dynamics associated with the spring and the collective dynamics.
Such mixing will be strong if the characteristic scale of the
internal vibrations $\omega_v$ is comparable to the scale associated
with collective motion $\omega_r$.  As both of these scales are
parametrically of order $N_c^0$ there is no reason associated with
$1/N_c$ physics for them not to mix strongly and thereby  ruin the
rigid rotor dynamics. This is the semiclassical argument outlined in
ref.~\cite{Coh04}. Underlying this is the simple time scale argument
of ref.~\cite{Coh03}.

An alternative way to see this is simply to look at the classical
dynamics of the system discussed in the appendix.  The essential
point there is to note that although there are two center of mass
degrees of freedom (say, moving in the $x$ and $y$ directions from
the north pole) there is  only {\it one} zero mode and it is
purely static. Rigid-rotor quantization is only legitimate for
true collective motion associated with zero modes.  Since there
is no dynamical zero mode in the problem, one does not expect
rigid-rotor quantization to apply.  It is worth noting here that
the analogous thing occurs for the chiral soliton case: it is
precisely the lack of dynamical zero modes which accounts for the
difference between the Callan-Klebanov and  rigid-rotor
approaches.

Of course, the model considered here is just a toy.  The real
question of interest is what form of quantization is correct for the
soliton models.  However, the analogy with the actual models is, in
fact, very close:  the key issues of topology, velocity-dependent
forces and the existence of collective and vibrational modes with
the same quantum numbers are present in both the toy problem and in
the problem of physical interest. Thus, this calculation should be
viewed as strong evidence that in the physically interesting
problems arising in soliton models the Callan-Klebanov approach is
correct at large $N_c$ and the rigid-rotor approach is generally
incorrect.

There are models for which the rigid rotor approximation does
work, such as the model considered in ref.~\cite{DP}. However, in
that case it is easy to see that the Callan-Klebanov approach
gives the same result as the rigid rotor method at large
$N_c$.\cite{Pob} Thus, the present situation is one in which the
Callan-Klebanov approach has been shown to be valid at large
$N_c$ for all cases studied, while the rigid rotor method is only
valid when it agrees with the Callan-Klebanov approach. This
strongly suggests that where the two methods disagree the
Callan-Klebanov quantization is the correct one. Since the two
approaches are known to disagree for chiral soliton
models\cite{IKOR}, it seems highly unlikely that the rigid rotor
quantization is valid; at large $N_c$ Callan-Klebanov
quantization is the correct approach. This strongly suggests that
the many calculations done using rigid rotor quantization for
exotic states of $SU(3)$ solitons are unjustified.

\acknowledgments The authors acknowledge constructive comments from
I.~ Klebanov, D.~Diakonov, D.~Dakin and P.~V.~Pobylitsa.  This work
was supported by the U.S.~Department of Energy through grant
DE-FG02-93ER-40762. One of the authors (AN) acknowledges the support
of the University of Maryland through the Senior Summer Scholars
program.

\section{Appendix}\label{append}
\begin{figure}
\label{threeOmegas}
  \begin{minipage}{2.5in}
    \includegraphics{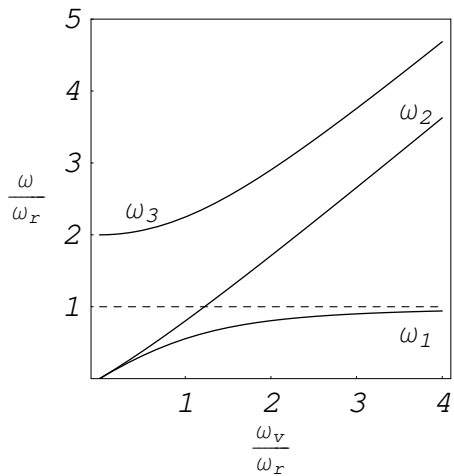}
    \caption{We plot the non-zero semiclassical eigenmodes of the
        system, $\omega_1$, $\omega_2$, $\omega_3$.}
  \end{minipage}
\end{figure}

Here we review the semiclassical treatment of the toy problem,
following ref.~\cite{Coh04}.  The first step is the treatment of
classical motion at small amplitude centered on the north pole.
For small amplitude motion, the system looks like a particle on a
plane with a constant perpendicular magnetic field. In this
regime, we can write the linearized equations of motion  for the
particles and their velocities to first order as
\begin{equation}\label{eom}
    \frac{d \mathbf{v}}{d t} = i \overline{M} \mathbf{v}
\end{equation} with
\begin{widetext}
\begin{equation}
\overline{M}= -i \left( \begin{array}{c c c c c c c c}
 0 & 0 & 0 &  0 & 1 & 0 & 0 & 0 \\
 0 & 0 & 0 &  0 & 0 & 1 & 0 & 0 \\
 0 & 0 & 0 &  0 & 0 & 0 & 1 & 0 \\
 0 & 0 & 0 &  0 & 0 & 0 & 0 & 1 \\
 \frac{-\omega_v^2}{2} & 0 & \frac{\omega_v^2}{2}& 0 & 0 & 2 \omega_r & 0 & 0 \\
 0 & \frac{-\omega_v^2}{2} & 0 & \frac{\omega_v^2}{2} &  -2\omega_r & 0 & 0 & 0  \\
\frac{\omega_v^2}{2} & 0 & \frac{-\omega_v^2}{2}& 0 & 0 & 0 & 0 & 0 \\
 0 & \frac{\omega_v^2}{2} & 0 & \frac{-\omega_v^2}{2} & 0 & 0 & 0 & 0  \end{array} \right)\; \mathrm{and} \;
\mathbf{v}= \left( \begin{array}{c} x_q\\ y_q\\ x_n \\
y_n\\\dot{x}_q\\\dot{y}_q\\\dot{x}_n\\\dot{y}_n \end{array} \right)
\; \; \; {\rm with} \; \; \; \omega_v=\sqrt{\frac{2 k}{m}} \; \;, \;
\; \omega_r=\frac{g q}{2 m R^2} \; . \label{m}\end{equation}
\end{widetext}
The subscripts $q$ and $n$ refer to the charged and neutral
particles respectively, and $x$ and $y$ refer to the Cartesian
coordinates.  Note that since there are velocity-dependent terms
it is natural to work in terms of coupled first-order differential
equations.  The canceling factors of $i$ in Eqs.~(\ref{eom}) and
(\ref{m}) are put in for later convenience.

We are interested in finding the eigenmodes of the system; {\it
i.e.}, harmonic solutions of the form $\mathbf{v}(t) = \mathbf{v}_j
\exp(-i \omega_j t)$.  Inserting this ansatz into Eq.~(\ref{eom})
yields a simple eigenvalue equation:
\begin{equation}\label{modeEq}
    \omega_j \mathbf{v}_j = \overline{M} \mathbf{v}_j \; .
\end{equation}
On physical grounds, we know the $\omega_j$ are real.  Moreover the
matrix $\overline{M}$ is purely imaginary which implies that if
${\mathbf v}_j$ is an eigenvector with eigenvalue $\omega_j$ then
${\mathbf v}_j^*$ is an eigenvector with eigenvalue $-\omega_j$.
Thus, the eigenvectors form pairs associated with positive and
negative frequency solutions.  We refer to one of these pairs
together an an eigenmode.

One of these modes is a zero frequency mode:
\begin{equation}
\mathbf{v}_0= \left( \begin{array}{c} 1\\ i \\ 1 \\ i \\
0\\ 0 \\ 0 \\ 0 \end{array} \right ) \;\;.
\end{equation}
A striking feature of this zero mode is the absence of any time
dependence in it: the four lower components associated with the
time derivatives are zero.  Thus, this mode is associated with
pure static rotations.  By extracting the real and imaginary
parts ($\mathbf{v}_0$ and $\mathbf{v}_0^*$ are degenerate since
$\omega=0$ and hence one can form linear combinations of the two)
we see that this mode corresponds to a collective
time-independent rotation of the two particles in either the $x$
or $y$ directions.  As this mode is completely non-dynamical,
when quantized it is associated with ``excitations'' which move
from one of the highly degenerate ground states of the theory to
another. In this context it is useful to recall that the
degeneracy of the ground state is $2 qg +1$ which diverges at
large $N_c$.

It is extremely important to note that the one zero mode found
above is the {\it only} zero mode for the system. We can solve Eq.
(\ref{modeEq}) to find the three non-zero frequency eigenmodes of
the system. The three eigenvalues can be computed analytically,
but the form of the solutions is cumbersome and not particularly
illuminating. We denote the three eigenfrequencies as $\omega_1$,
$\omega_2$, and $\omega_3$ defined such that $\omega_3 > \omega_2
> \omega_1$. These are plotted in terms of convenient
combinations of variables in Fig.~3.

As these modes are harmonic in the large $N_c$ limit, they can be
quantized trivially yielding Eq.~(\ref{n1n2n3}).

% ----------------------------------------------------------------
\bibliographystyle{amsplain}

\end{document}